\documentclass[sn-apa]{sn-jnl}% APA Reference Style
\jyear{2021}%

\theoremstyle{thmstyleone}%
\newtheorem{theorem}{Theorem}%  meant for continuous numbers
\theoremstyle{thmstyletwo}%
\newtheorem{remark}{Remark}%

\theoremstyle{thmstylethree}%
\usepackage{comment}
\usepackage[inline]{enumitem}
\usepackage{amsmath}
\usepackage{bm}
\usepackage{graphics}
\usepackage{multirow}
\usepackage{chngpage}
\usepackage{bbm}
\usepackage{multicol}

\usepackage{float}
\usepackage{anyfontsize}
\usepackage{lmodern}

\begin{document}

\title[Visualization for departures from symmetry with the divergence-type measure]{Visualization for departures from symmetry with the power-divergence-type measure in two-way contingency tables}


\author*[1]{\fnm{Wataru} \sur{Urasaki}}\email{urasaki.stat@gmail.com}
\author[2]{\fnm{Tomoyuki} \sur{Nakagawa}}\email{tomoyuki.nakagawa@meisei-u.ac.jp}
\equalcont{These authors contributed equally to this work.}
\author[3]{\fnm{Jun} \sur{Tsuchida}}\email{tsuchidj@kyoto-wu.ac.jp}
\equalcont{These authors contributed equally to this work.}
\author[1]{\fnm{Kouji} \sur{Tahata}}\email{kouji\_tahata@is.noda.tus.ac.jp}
\equalcont{These authors contributed equally to this work.}

\affil*[1]{\orgdiv{Department of Information Sciences}, \orgname{Tokyo University of Science}, \orgaddress{\city{Noda City}, \postcode{278-8510}, \state{Chiba}, \country{Japan}}}

\affil[2]{\orgdiv{School of Data Science}, \orgname{Meisei University}, \orgaddress{\city{Hino City}, \postcode{191-8506}, \state{Tokyo}, \country{Japan}}}

\affil[3]{\orgdiv{Department of Data Science}, \orgname{Kyoto Women's University}, \orgaddress{\city{Kyoto City}, \postcode{605-8501}, \state{Kyoto}, \country{Japan}}}


\abstract{
When the row and column variables consist of the same category in a two-way contingency table, it is specifically called a square contingency table.
Since it is clear that the square contingency tables have an association structure, a primary objective is to examine symmetric relationships and transitions between variables.
While various models and measures have been proposed to analyze these structures understanding changes between two variables in behavior at two-time points or cohorts, it is also necessary to require a detailed investigation of individual categories and their interrelationships, such as shifts in brand preferences.
This paper proposes a novel approach to correspondence analysis (CA) for evaluating departures from symmetry in square contingency tables with nominal categories, using a power-divergence-type measure.
The approach ensures that well-known divergences can also be visualized and, regardless of the divergence used, the CA plot consists of two principal axes with equal contribution rates.
Additionally, the scaling is independent of sample size, making it well-suited for comparing departures from symmetry across multiple contingency tables.
Confidence regions are also constructed to enhance the accuracy of the CA plot.
}





\maketitle

%\clearpage
%%%%%%%%%%%%%%%%%%%%%%%%%%%%%%%%%%%%%%%%%%%%%%%%%%%%%%%%%%%%%%%%%%%%%%%%%%
\section{Introduction}\label{sec1}
The same categorical variable appears in various fields, including medicine, education, and social science, and has long been the subject of analysis.
Specifically, a table derived from the combination of the same categorical variables is known as a square contingency table.
In a widely accepted analysis of contingency tables, numerous studies have evaluated the independence or association between variables, and recently, new methods have been proposed by \cite{kateri2018phi, forcina2021new, chatterjee2021new, urasaki2023generalized}.
However, in the square contingency table, the observed values are concentrated in the diagonal components and tend to decrease as they move away from the diagonal components.
The feature clearly shows a strong association, making traditional analysis methods inappropriate.
Therefore, research focusing on symmetry and the transition between variables has advanced in the context of square contingency table analysis.
%For such studies, it is interesting to investigate how similar or transitional, the variable is between the two-time points or cohorts and so a study of the departure from symmetry of the variables is important.
It is important to investigate how similar or transitional the variables are between the two-time points or cohorts, and exploring the departure from symmetry is also of interest.

%\cite{bowker1948test} seems to be the first study on symmetry in history, and various proposals have been made to date.
There are two well-known methodological approaches to the analysis of symmetry: ``Model'' and ``Measure''.
As for models, in addition to the symmetry (S) model (\citealp{bowker1948test}), the marginal homogeneity (MH) model (\citealp{stuart1955test}), and the quasi symmetry (QS) model (\citealp{caussinus1965contribution}) are well known, and recently proposed models with divergence by \cite{kateri2021families, tahata2022advances}.
%Additionally, asymmetry models such as the conditional symmetry (CS) model (\citealp{mccullagh1978class}), the diagonals-parameter symmetry (DPS) model (\citealp{goodman1979multiplicative}), and the linear diagonals-parameter symmetry (LDPS) model (\citealp{tomizawa1993diagonals}) and so on, also exist for symmetric categories with transitional structures between the two-time points or cohorts.
On the other hand, the measure is a method to evaluate the degree of departure from the models within a fixed interval regardless of sample size.
These features enable the quantification and comparison of the degrees of departure from the model for each contingency table observed across various factors, including confounding.
As an example, \cite{tomizawa1998power} proposed a generalized divergence-type measure that guarantees that the degree of departure from the S model can be evaluated in the range of 0 to 1 for divergences included in power-divergence.
(For more detail on the divergence and power-divergence, see \citealp{renyi1961measures, cressie1984multinomial, read1988goodness}.)

While many methods have been proposed to understand the overall structure of contingency tables, others have been proposed for a long time to understand the relationships among categories through visualization.
The method is called ``Correspondence Analysis (CA)'' proposed by \cite{benzecri1973analyse}, and sophisticated and easy-to-understand visualizations have been realized, enabling quick interpretation and understanding of data.
In particular, simple CA is still widely used and easily implemented using tools such as SAS, R, and Python.
Simple CA is a visualization based on Pearson's chi-square statistic, which is described in detail for \cite{greenacre2017correspondence, beh2014correspondence}.
Additionally, visualization based on the Goodman-Kruskal's tau index, which is one of the association measures, and the approximations of divergence statistics have been proposed by \cite{beh2010non, beh2023correspondence}.
Many proposals aim to understand relationships among categories of the contingency table based on independence or association evaluation methods.
Conversely, although visualizing departures from symmetry has been proposed by \cite{beh2022visualising, greenacre2000correspondence, constantine1978graphical, beh2024correspondence}, all are based on divergence statistics.

This paper proposes a methodological approach to visually provide the relationship between nominal categories based on the degree of departure from symmetry by the power-divergence-type measure in two-way square contingency tables.
%The proposal guarantees that several well-known divergences, including Pearson divergence and KL-divergence, can be visualized, and shows the advantages of using a power-divergence-type measure.
This proposal ensures that several important divergence-type measures, including Hellinger distance, Kullback–Leibler (KL) divergence, Cressie-Read divergence, and Pearson divergence, can also be visualized.
In the visualization based on our proposal, the same category of row and column variables can show the same degree of departure on a CA plot without depending on the divergence's parameters.
Section \ref{sec5} introduces the construction of confidence regions due to concerns about the accuracy of depicting the degree of departure from symmetry for each category by a single point on the CA plot. 
In Section \ref{sec6}, by using the proposed method for real data, we discuss the different views given by various divergences and the choice of parameters.
Additionally, Section \ref{sec7} presents a discussion on the reasons for employing the power-divergence-type measure, and Section \ref{sec8} provides the conclusion.
These may provide new insights into symmetry.

%%%%%%%%%%%%%%%%%%%%%%%%%%%%%%%%%%%%%%%%%%%%%%%%%%%%%%%%%%%%%%%%%%%%%%%%%%
\section{Analysis of Symmetry}\label{sec2}
Consider an $R\times R$ contingency table with nominal categories for the row variable $X$ and the column variable $Y$.
Let $p_{ij}$ denote the probability that an observation will fall in the $i$th row and $j$th column of the table ($i=1,\ldots, R;j=1,\ldots, R)$.
Additionally, let $p_{i\cdot}$ and $p_{\cdot j}$ be denoted as $p_{i\cdot}=\sum_{j=1}^R p_{ij}$ and $p_{\cdot j}=\sum_{i=1}^R p_{ij}$.
Conversely, let $n_{ij}$ donate the observed frequency in the $i$th row and $j$th column of the table.
The totals $n_{i\cdot}$, $n_{\cdot j}$ and $n$ are also denoted as $n_{i\cdot}=\sum_{j=1}^R n_{ij}$, $n_{\cdot j}=\sum_{i=1}^R n_{ij}$, and $n = \sum\sum n_{ij}$, respectively.
Assuming a multinomial distribution for the $R\times R$ contingency table, a maximum likelihood estimator of $p_{ij}$, $\hat{p}_{ij}$, is express as $\hat{p}_{ij} = n_{ij}/n$.
Using this notation, we introduce Bowker's test and power-divergence-type measure in the analysis of symmetry.

%%%%%%%%%%%%%%%%%%%%%%%%%%%%%%%%%%%%%%%%%%%
\subsection{Bowker's test}\label{sec2.1}
Bowker's test, which is a test of symmetry, may be undertaken by defining the S model with
\begin{align*}
H_0 \: : \: p_{ij} = p_{ji}; \; \forall i, j.
\end{align*}
Under the null hypothesis $H_0$, Bowker's $\chi^2$ statistics is given as follows:
\begin{align*}
\chi^2_{S} &= \frac{1}{2}\sum^R_{i=1}\sum^R_{j=1}\frac{(n_{ij}-n_{ji})^2}{n_{ij}+ n_{ji}} \\
&= \mathop{\sum \sum}_{i < j}\frac{(n_{ij}-n_{ji})^2}{n_{ij}+n_{ji}},
\end{align*}
and follows a chi-squared distribution with $R(R-1)/2$ degrees of freedom.
$H_0$ is rejected for high values of $\chi^2_{S}$.
Bowker's test is a generalization of McNemar's test (\citealp{mcnemar1947note}) for $R\times R$ contingency table with $R>2$.
\cite{beh2022visualising} proposed visualization of the degree of departures from symmetry based on $\chi^2_S/n$.

%%%%%%%%%%%%%%%%%%%%%%%%%%%%%%%%%%%%%%%%%%%
\subsection{Power-divergence-type Measure}\label{sec2.2}
When the S model does not hold by Bowker's test, one interest is to quantitatively evaluate the degree of departure from symmetry.
\cite{tomizawa1998power} proposed the following power-divergence-type measure:
\begin{align*}
\Phi^{(\lambda)} = \frac{\lambda(\lambda+1)}{2^\lambda-1}I^{(\lambda)}\left(\{p^{*}_{ij};p^{s}_{ij}\} \right), \quad \lambda > -1
\end{align*}
where
\begin{align*}
I^{(\lambda)}\left(\{p^{*}_{ij};p^{s}_{ij}\} \right) = \frac{1}{\lambda(\lambda+1)} \mathop{\sum \sum}_{i \neq j}p^{*}_{ij}\left[ \left( \frac{p^{*}_{ij}}{p^{s}_{ij}}\right)^{\lambda} - 1 \right].
\end{align*}
Assume the $\{p_{ij}+p_{ji}\}$ for $i\neq j$ are all positive, and $p^{*}_{ij}$ and $p^s_{ij}$ are defined as $p^*_{ij}=p_{ij}/\delta$ and $p^s_{ij}=(p^*_{ij} + p^*_{ji})/2$, with $\delta = \mathop{\sum \sum}_{i \neq j} p_{ij}$.
The measure $\Phi^{(\lambda)}$ has the three properties: 
\begin{theorem}
The measure $\Phi^{(\lambda)}$ satisfies the following properties for all $\lambda$:
\begin{itemize}
\item[1] $\Phi^{(\lambda)}$ must lie between 0 and 1.
\item[2] $\Phi^{(\lambda)} = 0$ if and only if there is a complete structure of symmetry, i.e., $p_{ij}=p_{ji}$.
% for all $i<j$ っている？
\item[3] $\Phi^{(\lambda)} = 1$ if and only if there is a structure in which the degree of departure of symmetry is the largest, i.e., $p_{ij} = 0$ (then $p_{ji} > 0$) or $p_{ji} = 0$ (then $p_{ij} > 0$).
\end{itemize}
\end{theorem}
The parameter $\lambda$ is determined by the user, and the limit as $\lambda \rightarrow 0$ is taken for the value at $\lambda = 0$.
In particular, when $\lambda = - 1/2$, $0$, $2/3$, and $1$, it is a famous divergence with a special name, called Hellinger distance, KL divergence, Cressie-Read divergence, and Pearson divergence, respectively.
When analyzing real data using the measure, the estimated values of $\Phi^{(\lambda)}$ are calculated using the plug-in estimators $p^{*}_{ij}$ and $p^{s}_{ij}$, where $p_{ij}$ is replaced by $\hat{p}_{ij}$.
The important point here is that the sample size $n$ does not appear in the estimator of $\Phi^{(\lambda)}$.
Therefore, the major benefit of the measure is that it can be quantified independently of the sample size, making it suitable for comparing the asymmetry of multiple contingency tables.

%%%%%%%%%%%%%%%%%%%%%%%%%%%%%%%%%%%%%%%%%%%%%%%%%%%%%%%%%%%%%%%%%%%%%%%%%%
\section{Visualization and Power-divergence-type Measure}\label{sec3}
Visualization of relationships between categories in contingency tables enables rapid interpretation and understanding of data, even for non-experts.
In particular, when examining the symmetry of nominal categories, it is important to analyze the similar or transitional relationship of categories between the two-time points or cohorts.
This section shows that the power-divergence-type measure $\Phi^{(\lambda)}$ can be used to visualize the symmetric structures and interrelationships of individual categories.

%%%%%%%%%%%%%%%%%%%%%%%%%%%%%%%%%%%%%%%%%%%
\subsection{Simple CA and Power-divergence-type Measure}\label{sec3.1}
The power-divergence-type measure $\Phi^{(\lambda)}$ can also be defined as:
\begin{align*}
\Phi^{(\lambda)} &= \mathop{\sum \sum}_{i \neq j}\frac{p^{*}_{ij}+p^{*}_{ji}}{2}\left[1- \frac{\lambda 2^\lambda}{2^\lambda-1}H^{(\lambda)}_{ij}(\{p^{c}_{ij},p^{c}_{ji}\}) \right] =  \mathop{\sum \sum}_{i \neq j} \phi^{(\lambda)}_{ij},
\end{align*}
where 
\begin{align*}
H^{(\lambda)}_{ij}(\{p^{c}_{ij},p^{c}_{ji}\}) &= \frac{1}{\lambda} \left[1-(p^{c}_{ij})^{\lambda+1} - (p^{c}_{ji})^{\lambda+1} \right], \\
\phi^{(\lambda)}_{ij} &= \frac{p^{*}_{ij}+p^{*}_{ji}}{2}\left[1- \frac{\lambda 2^\lambda}{2^\lambda-1}H^{(\lambda)}_{ij}(\{p^{c}_{ij},p^{c}_{ji}\}) \right],
\end{align*}
and $p^{c}_{ij} = p_{ij}/(p_{ij}+p_{ji})$.
$\phi^{(\lambda)}_{ij}$ represents the departure from symmetry for each ($i, j$) cell of the contingency table and is non-negative.
Let consider the $S\times S$ matrix $S_{skew(\lambda)}$ with the following ($i, j$) elements:
\begin{align*}
s_{ij} = \text{sign}(p_{ij}-p_{ji}) \sqrt{\phi^{(\lambda)}_{ij}}, 
%\begin{cases}
%\;\;\; \sqrt{\phi^{(\lambda)}_{ij}} &(i<j) \\
%-\sqrt{\phi^{(\lambda)}_{ij}} &(i>j).
%\end{cases}
\end{align*}
where $\text{sign}(x)$ is a sign function defined as follows
\begin{equation*}
\text{sign}(x) = \left\{
\begin{array}{rl}
1 & (x > 0),\\
0 & (x = 0), \\
-1 & ( x < 0).
\end{array}
\right.
\end{equation*}
The matrix with such elements is called an anti-symmetric or skew-symmetric matrix, and $S_{skew(\lambda)}^T=-S_{skew(\lambda)}$.
Using this matrix, the measure $\Phi^{(\lambda)}$ can be expressed as
\begin{align*}
\Phi^{(\lambda)} &= \mathop{\sum \sum}_{i \neq j} \phi^{(\lambda)}_{ij} \\
&= trace(S_{skew(\lambda)}^T S_{skew(\lambda)}) \\
&= trace(S_{skew(\lambda)} S_{skew(\lambda)}^T).
\end{align*}
It indicates that $S_{skew(\lambda)}$ can reconstruct the measure $\Phi^{(\lambda)}$ while having information about the degree of departure from symmetry for each cell.

To visualize departures from symmetry, the matrices represented the principal coordinates of the row and column categories are obtained from the singular value decomposition (SVD) of $S_{skew(\lambda)}$, that is, 
\begin{align*}
S_{skew(\lambda)} &= AD_{\mu}B^T,
\end{align*}
where $A$ and $B$ are $R\times M$ orthogonal matrices containing left and right singular vectors of $S_{skew(\lambda)}$, respectively, and $D_\mu$ is $M\times M$ diagonal matrix with singular values $\mu_m$ ($m =1,\dots, M$).
The singular values $\mu_m$ are also lined up in consecutive pairs of values, so that  $1>\mu_1 = \mu_2 > \mu_3 = \mu_4 > \dots \geq 0$.
Since $S_{skew(\lambda)}$ is a skew-symmetric matrix, $M$ varies depending on the number of categories.
Therefore,
\begin{align*}
M =
\begin{cases}
R, &\text{$R$ is even}, \\
R-1, &\text{$R$ is odd}.
\end{cases}
\end{align*}
For a detailed description of features of the skew-symmetric matrix, see \cite{ward1978eigensystem, gower1977analysis}.
To obtain the principal coordinates of row and column categories using these matrices $A$, $B$, and $D_\mu$ in the case of symmetry, it is necessary to consider metrics in a different approach from traditional simple CA for independence.
In the case of independence, the principal coordinates are obtained from diagonal matrices $D_r$ and $D_c$ with $p_{i\cdot}$ and $p_{\cdot j}$, respectively, assuming that each row and column variable has a different metric.
Conversely, in the case of symmetry, the row and column variables are assumed to have the same metric, so we use the metric $D=(D_r + D_c)/2$ adopted by \cite{greenacre2000correspondence}.
Therefore, a visual representation of the departure from symmetry can be achieved by plotting the principal coordinates of the row and column categories defined as 
\begin{align*}
\begin{cases}
F = D^{-1/2}AD_{\mu}, \\
G = D^{-1/2}BD_{\mu},
\end{cases}
\end{align*}
respectively.

From \cite{gower1977analysis}, the SVD of the skew-symmetric matrix $S_{skew (\lambda)}$ can also be expressed as:
\begin{align*}
S_{skew(\lambda)} &= AD_{\mu}J_M A^T,
\end{align*}
where
\begin{align*}
B &= AJ_M^T.
\end{align*}
And $J_M$ is a block diagonal and orthogonal skew-symmetric matrix.
In particular, when $R=2$, 
\begin{align*}
J_2 &= 
\begin{pmatrix}
0 & 1 \\
-1 & 0 
\end{pmatrix}
,
\end{align*}
and when $R$ is even,
\begin{align*}
J_R &= 
\begin{pmatrix}
J_{R-2} & O_2 \\
O_2 & J_2 
\end{pmatrix}
,
\end{align*}
where $O_2$ is a $2 \times 2$ zero matrix.
Conversely, if $R$ is odd, the ($R, R$)th element is 1, and the other $R$th row and column elements are 0, that is, 
\begin{align*}
J_R &= 
\begin{pmatrix}
J_{R-1} & \bm{0}^T_{R-1} \\
\bm{0}_{R-1} & 1
\end{pmatrix}
,
\end{align*}
where $\bm{0}_{R-1}$ is a zero vector of length $R-1$.
Therefore, column coordinates can be expressed in row coordinates, where 
\begin{align*}
G %&= D^{-1/2}AJ_M^TD_{\mu} \\
%&= D^{-1/2}AD_{\mu}J_M^T \\
&= FJ_M^T.
\end{align*}
%Similarly, for row coordinates, $F=GJ_M^T$ can be derived, so that plots of the same category of rows and columns are rotated around the origin.
Similarly, since $F=GJ_M$ can be derived for row coordinates, the positions of row and column coordinates in the same category coincide when rotated around the origin.

%%%%%%%%%%%%%%%%%%%%%%%%%%%%%%%%%%%%%%%%%%%
\subsection{Total Inertia}\label{sec3.2}
When visualizing with the principal coordinates, it is possible to quantify how much the coordinate axes in $m$ dimensions ($m=1,\dots, M$) reflect the degree of departure from symmetry by calculating the total inertia of $S_{skew (\lambda)}$.
The total inertia can be expressed as the sum of squares of the singular values, that is,
\begin{align*}
\Phi^{(\lambda)} &= trace(S_{skew(\lambda)}^T S_{skew(\lambda)}) \\
&= trace((AD_{\mu}B^T)^T (AD_{\mu}B^T)) \\
&= trace(D_{\mu}^2).
\end{align*}
When constructing the CA plot, the principal inertia of the $m$th dimension is assumed to be $\mu_m$, so that visualization up to the maximum $M$ dimensions can be realized.
The fact that the total inertia can be expressed as the sum of squares of the singular values determines the contribution ratio that indicates how much the coordinate axes of each dimension are reflected from the values of $\Phi^{(\lambda)}$.
Therefore, the contribution ratio of the $m$th dimension is calculated by 
\begin{align*}
100 \times \frac{\mu_{m}^2}{\Phi^{(\lambda)}}.
\end{align*}
The CA plot of up to $M$ dimensions can be constructed, but visualization of up to two dimensions is preferable due to the limitations of human visual cognitive ability.
Because of the limitation, it is necessary to select any two dimensions, but in this case, it is appropriate to use the first and second dimensions.
The singular values obtained from the skew-symmetric matrix are such that $\mu_1$ and $\mu_2$ are the largest pairs, so the first and second dimensions are the best visually optimal representations reflecting the departures from symmetry.
And since $\mu_1 =\mu_2$, the two dimensions can be equally represented.

Total inertia can also be derived from row or column principal coordinates.
For example, in the case of row coordinates,
\begin{align*}
trace(F^TDF) &= trace((D^{-1/2}AD_{\mu})^T D (D^{-1/2}AD_{\mu})) \\
&= trace(D_{\mu}^2) \\
&= \Phi^{(\lambda)}.
\end{align*}
It indicates that the coordinates of all row categories match the origin if $\Phi^{(\lambda)} = 0$, indicating that the square contingency table has a complete structure of symmetry.
%Therefore, it can be seen that the more departures from symmetry are observed for each category, the more they are plotted away from the origin.
The same can be said for the column coordinates from 
\begin{align*}
trace(G^TDG) &= \Phi^{(\lambda)}.
\end{align*}
When the square contingency table has a complete structure of symmetry, all row and column categories are plotted at the origin.
On the other hand, we consider the case where there are partial symmetric structures.
The principal coordinates of the row and column categories are also expressed as follows using $S_{skew(\lambda)}$:
\begin{align*}
\begin{cases}
F = D^{-1/2}S_{skew (\lambda)}B, \\
G = D^{-1/2}S_{skew (\lambda)}A.
\end{cases}
\end{align*}
If one considers the ($i, m$)th element $f_{im}$ of the row coordinate matrix $F$,
\begin{align*}
f_{im} &= \frac{\sqrt{2}}{\sqrt{p_{i\cdot} + p_{\cdot i}}}\sum^R_{j=1}s_{ij}b_{jm} 
\end{align*}
where $b_{jm}$ is the ($j,m$)th element of $B$.
Note that $s_{ij}$ reflects the ($i, j$)th cell's partial degree of departure from symmetry, $\phi_{ij}$.
Therefore, the fact that the $i$th category is located at the origin means that the $i$th row and column have a perfect symmetric structure, i.e., $p_{ij}=p_{ji}$ holds for all $j=1,\dots, R$.
The same is true for column coordinates.

%%%%%%%%%%%%%%%%%%%%%%%%%%%%%%%%%%%%%%%%%%%%%%%%%%%%%%%%%%%%%%%%%%%%%%%%%%
\section{Confidence Regions for Individual Categories}\label{sec5}
In Section \ref{sec3}, we demonstrated that visualization for departures from symmetry indicates complete symmetry when located at the origin, and the distance from the origin signifies the degree of departure from symmetry for each category.
It is evident that visualizing categories in terms of symmetry is important, but it is also crucial to understand the interrelationships among categories.
As for independence, \cite{Lebart1984multivariate, beh2001confidence, beh2010elliptical, beh2023confidence} proposed constructing a confidence circle and a confidence ellipse for each category given a significance level $\alpha$, based on the simple CA.
Additionally, \cite{ringrose1992bootstrapping, ringrose1996alternative, greenacre2017correspondence} proposed constructing a non-circular confidence region using a convex hull by applying a bootstrap method.
\cite{greenacre2017correspondence} also proposed a method for constructing confidence intervals using the delta method (see, \citealp{agresti2012categorical, bishop2007discrete}), but it also reported some problems.
In this section, we also discuss the construction of the confidence regions in symmetry.

Let $n_{ij}$ denote the observed frequency at the intersection of the $i$th row and $j$th column within the table.
Assuming a multinomial distribution applies to the $R \times R$ table, the sample version of $\Phi^{(\lambda)}$, denoted as $\hat{\Phi}^{(\lambda)}$, is derived from $\Phi^{(\lambda)}$ with $p_{ij}$ replaced by $\hat{p}_{ij}$.
The $100(1-\alpha)\%$ confidence region for the $i$th row category in the two-dimensional CA plot is expressed as follows
\begin{align*}
\frac{(x-f_{i1})^2}{x^2_{i(\alpha)}} + \frac{(y-f_{i2})^2}{y^2_{i(\alpha)}} = 1,
\end{align*}
where
\begin{align*}
x_{i(\alpha)} &= \left(\frac{\hat{p}_{i\cdot}+\hat{p}_{\cdot i}}{2}\right)^{-1/2}\mu_1 \sqrt{\frac{\chi^2_\alpha}{\hat{\Phi}^{(\lambda)}/\frac{\lambda(\lambda+1)}{2n\hat{\delta}(2^\lambda-1)}} \left(1-\sum^M_{m=3}a^2_{im}\right) }, \\
y_{i(\alpha)} &= \left(\frac{\hat{p}_{i\cdot}+\hat{p}_{\cdot i}}{2}\right)^{-1/2}\mu_2 \sqrt{\frac{\chi^2_\alpha}{\hat{\Phi}^{(\lambda)}/\frac{\lambda(\lambda+1)}{2n\hat{\delta}(2^\lambda-1)}} \left(1-\sum^M_{m=3}a^2_{im}\right)}.
\end{align*}
And $\hat{\delta}$ is the plug-in estimator with $p_{ij}$ replaced by $\hat{p}_{ij}$.
The $a_{im}$ is the ($i,m$)th element of $A$.
In addition, $\chi^2_\alpha$ is the upper $\alpha \%$ point of the chi-square distribution with $R(R-1)/2$ degrees of freedom.
The proof for constructing the confidence region is given in Appendix \ref{app:CR}.
%The derivation of the confidence region can be obtained in the same way as that proposed by \cite{beh2010elliptical}.
Note that the singular values $\mu_1$ and $\mu_2$ are equal, so the confidence regions are circular.
%It is due to the use of a skew-symmetric matrix, whereas in the case of independence, the confidence regions are elliptical.

%%%%%%%%%%%%%%%%%%%%%%%%%%%%%%%%%%%%%%%%%%%%%%%%%%%%%%%%%%%%%%%%%%%%%%%%%%
\section{Numerical Experiment}\label{sec6}
Consider Table \ref{coffee} by \cite{agresti2019introduction}, where the original data comes from \cite{grover1987simultaneous}. 
Table \ref{coffee} shows data on decaffeinated coffee purchases for the first and second time.
The symmetry of the row and column variables in such data implies that there is a balanced inflow of people who tend to choose different products regardless of the product they initially selected.
Therefore, departures from symmetry in each category indicate that for a given coffee brand, the number of new buyers and those who stopped buying are uneven.
It is important to investigate whether there is any difference in the inflow of people in each category and how categories' interrelationships there are, so analysis using our proposal is necessary.
For the analysis, $\lambda=-1/2$, $0$, $2/3$, and $1$ are applied to $\Phi^{(\lambda)}$ as parameters to visualize using divergences that are significant enough to be given a special name.
Based on the relationship between row and column coordinates shown in Section \ref{sec3}, only row categories were plotted.

\begin{table}[h]
\caption{Choice of first and second purchases of five brands of decaffeinated coffee}
\centering
\label{coffee}
\begin{tabular}{rcccccc}
\hline
 & \multicolumn{5}{c}{Second Purchase} & \\ \cline{2-6}
\multicolumn{1}{c}{First Purchase} & High Pt & Taster's & Sanka & Nescaf\'{e} & Brim & Total  \\ 
%\multicolumn{1}{c}{First Purchase} & (hp) & (tc) & (sa) & (ne) & (br) & Total  \\ 
\hline
High Point (HP) & 93 & 17 & 44 & 7 & 10 & 171  \\
Taster's Choice (TC) & 9 & 46 & 11 & 0 & 9 & 75 \\
Sanka (SA) & 17 & 11 & 155 & 9 & 12 & 204 \\
Nescaf\'{e} (NE) & 6 & 4 & 9 & 15 & 2 & 36 \\
Brim (BR)  & 10 & 4 & 12 & 2 & 27 & 55 \\   \hline
Total & 135 & 82 & 231 & 33 & 60 & 541 \\ \hline
%\multicolumn{7}{l}{Source : Grover and Srinivasan (1987)}  \\
\end{tabular}
\end{table}

\begin{figure}[htbp]
\begin{tabular}{ccc}
\begin{minipage}{.47\textwidth}
\centering
\includegraphics[width=1\linewidth]{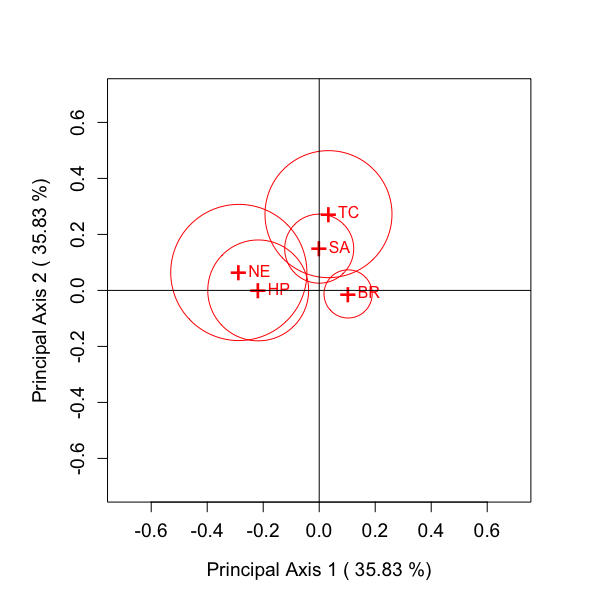}
\caption{$\lambda=-1/2$ (Hellinger distance)}
\label{f5}
\end{minipage}
\begin{minipage}{.47\textwidth}
\centering
\includegraphics[width=1\linewidth]{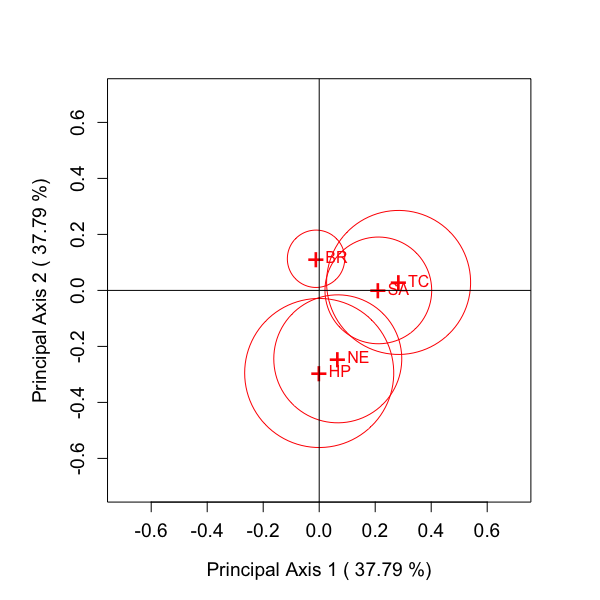}
\caption{$\lambda=0$ (KL divergence)}
\label{f6}
\end{minipage}
\end{tabular}

\begin{tabular}{ccc}
\begin{minipage}{.47\textwidth}
\centering
\includegraphics[width=1\linewidth]{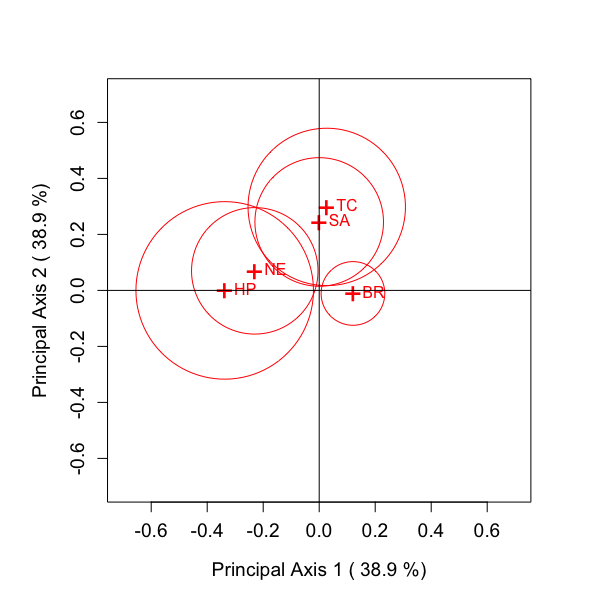}
\caption{$\lambda=2/3$ (Cressie-Read divergence)}
\label{f7}
\end{minipage}
\begin{minipage}{.47\textwidth}
\centering
\includegraphics[width=1\linewidth]{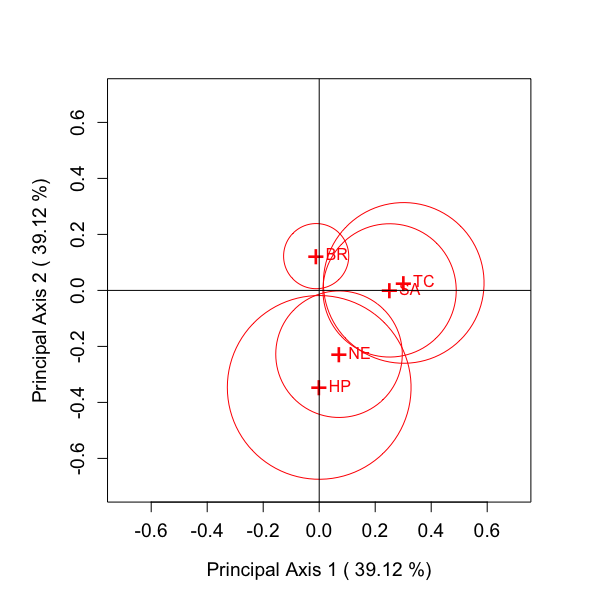}
\caption{$\lambda=1$ (Pearson divergence)}
\label{f8}
\end{minipage}
\end{tabular}
\end{figure}

Figures \ref{f5}-\ref{f8} show the results of analysis with each parameter.
As can be seen from these figures, ``Brim'' is located close to the origin for all parameters, while ``High Point'' is located far from the origin for many parameters.
Other brands can also be seen to be located far from the origin.
Therefore, the following can be considered for each brand.
\begin{itemize}
\item ``Brim'' has a strong symmetry, indicating that there are few differences between the first and second purchase choices.
Therefore, we can judge that Brim's decaffeinated coffee is selling steadily.
\item ``High Point'' has a larger departure from symmetry than other brands, indicating a considerable difference in the number of buyers on the first and second purchases.
As seen from the data, the number of second-time buyers is decreasing, so it can be concluded that there was some kind of mismatch and they moved to another brand.
\item For ``Taster's Choice'', ``Nescaf{\'e}'' and ``Sanka'', it can be seen that there have been some changes in the purchase choices.
\end{itemize}
Additionally, we can also confirm that the confidence regions of all brands do not include the origin for each parameter.
It indicates that, regardless of the differences in measurement methods for each famous divergence, there are significant departures from symmetry across all brands with sufficient accuracy.

Next, we focus on the positioning of the plot.
For each parameter, it can be observed that ``Taster's Choice'', ``Sanka'', and ``Brim'', which are located in the positive direction of principle axis 1 or 2, show an increase in the total number of purchases in the second selection compared to the first.
On the other hand, ``High Point'' and ``Nescaf{\'e}'', positioned in the negative direction, show a decrease in purchases.
Therefore, these results suggest that the primary axes reflect the overall changes in the number of buyers for each brand.
Another important observation is the pairs of brands that are in orthogonal positions.
As seen in the figures, the pairs of ``High Point'' and ``Sanka'', as well as ``Taster's Choice'' and ``Brim'', are confirmed to be orthogonal. 
In the context of visualizing the departures from symmetry, orthogonality represents a different meaning than independence; it indicates that one category in the pair only transitioned from the other category.
For example, Table \ref{coffee} shows that ``Sanka'' exhibits no customer shift with respect to ``High Point'', and similarly, ``Brim'' shows no change in relation to ``Taster's Choice''.
Our proposed method effectively visualizes these relationships, as demonstrated in this analysis.

While these insights can be easily derived from simply examining the table, the strength of our method lies in its ability to convey the same conclusions at a glance through the plot, regardless of the parameters. 
This advantage becomes more pronounced as the number of categories increases, offering a concise evaluation of the symmetry relationships between categories.

\begin{remark}\label{remark}
(Brief guideline of parameter selection)
Although the results obtained using the parameters in the numerical experiment did not show, it is often the case that different parameters can lead to different conclusions.
Therefore, selecting the optimal parameters remains one of the challenges to be addressed in this study, but, no mathematically valid method has been proposed until now.
%, and parameter selection is one of the important tasks in this study as well.
One practical method for parameter selection is to determine the parameters based on the user's desired divergence, taking into account the background of the data.
Additionally, in the case of CA, the parameters can be changed by 0.01 and analyzed in detail, and the one that maximizes the total contribution ratio of the first and second axes can be selected.
This method is also introduced by \cite{beh2023correspondence, beh2024correspondence}.
However, even in such a method, it is unclear from what point of view the divergence given by the selected parameters evaluates the departures from symmetry.
In conclusion, the user should adopt the divergences which have properties suitable for the analytical purpose.
If not possible, it would be better to select several well-known divergences and examine them in an exploratory manner.
\end{remark}

%%%%%%%%%%%%%%%%%%%%%%%%%%%%%%%%%%%%%%%%%%%%%%%%%%%%%%%%%%%%%%%%%%%%%%%%%%
\section{Discussion : Why use a measure ?}\label{sec7}
In Section \ref{sec3}, we propose a novel approach to visualize the relationship between nominal categories using a power-divergence-based measure $\Phi^{(\lambda)}$, focusing on departures from symmetry. 
Unlike traditional simple CA, which visualizes based on test statistics, our method offers a different evaluating way of these departures. 
The main difference is whether the method is influenced by sample size.

Consider two $R \times R$ square contingency tables with sample sizes $n_1$ and $n_2$, respectively. 
In the method by \cite{beh2022visualising, beh2024correspondence}, if $\chi^2_{S(1)}$ and $\chi^2_{S(2)}$ are the test statistics for the two tables, CA plots of the two can be created based on $\chi^2_{S(1)}/n_1$ and $\chi^2_{S(2)}/n_2$. 
However, it is important to note that because the scaling differs by $n_1$ and $n_2$, comparing the two CA plots is not straightforward.
Therefore, when visualizing multiple contingency tables, special procedures such as Multiple CA or Joint CA must be used. 
For detailed explanations of Multiple CA and Joint CA, refer to works such as \cite{greenacre2017correspondence, beh2014correspondence}.
Of course, proposals for comparing and visualizing categories by analyzing sum and difference components of several tables have also been made, such as in \cite{greenacre2003singular}. 
However, \cite{greenacre2003singular} pointed out for CA that, unless the sample sizes match, the differences in the patterns of category characteristics between tables are primarily influenced by the differences in sample size.

Let's reconsider our approach using the power-divergence-type measure. 
As discussed in Section \ref{sec2.2}, the measure is independent of sample size, making it suitable for comparison. 
Therefore, by using the measure, we demonstrate that it is possible to visually analyze both the sum and difference components of the tables.

\subsection{Approach for Matched Square Contingency Tables}\label{sec7.1}
For the two $R \times R$ square contingency tables, let the skew-symmetric matrices constructed by our proposed method be $S_{1(\lambda)}$ and $S_{2(\lambda)}$. 
These two matrices are assumed to be constructed using the same value of parameter $\lambda$.
The SVDs of the sum $S_{+(\lambda)} = S_{1(\lambda)} + S_{2(\lambda)}$ and the difference $S_{-(\lambda)} = S_{1(\lambda)} - S_{2(\lambda)}$ can be derived from the SVD of the following block matrix:
\begin{align*}
S_{\pm(\lambda)} &= 
\begin{pmatrix}
S_{1(\lambda)} & S_{2(\lambda)} \\
S_{2(\lambda)} & S_{1(\lambda)} 
\end{pmatrix}
.
\end{align*}
Suppose that the SVDs of $S_{+(\lambda)}$ and $S_{-(\lambda)}$ are respectively
\begin{align*}
S_{+(\lambda)} &= A_+ D_{\mu+} B^T_+, \\
S_{-(\lambda)} &= A_- D_{\mu-} B^T_-,
\end{align*}
where $A_+$, $B_+$, $A_-$, and $B_-$ are $R \times M$ orthogonal matrices containing left and right singular vectors of $S_+$ and $S_-$, respectively.
Additionally, $D_{\mu+}$ and $D_{\mu-}$ are diagonal matrices of the singular values $\mu_{m+}$ and $\mu_{m-}$ ($m=1, \dots, M$) in each case.
Note that $S_{\pm(\lambda)}$, $S_{+(\lambda)}$, and $S_{-(\lambda)}$ also become skew-symmetric matrices.
Then the SVD of $S_{\pm(\lambda)}$ can be also expressed
\begin{align*}
S_{\pm(\lambda)} &= \frac{1}{\sqrt{2}}
\begin{pmatrix}
A_+ & A_- \\
A_+ & -A_-
\end{pmatrix}
\begin{pmatrix}
D_{\mu+} & O \\
O & D_{\mu-}
\end{pmatrix}
\frac{1}{\sqrt{2}}
\begin{pmatrix}
B_+ & B_- \\
B_+ & -B_-
\end{pmatrix}
^T.
\end{align*}
The details of the SVD can be found in reference \cite{greenacre2003singular}.
One notable point to keep in mind when performing the SVD of this block matrix is that the matrices obtained from the SVDs of $S_{+(\lambda)}$ and $S_{-(\lambda)}$ do not appear separated.
Instead, they are interleaved according to the descending order of the values of $\mu_{m+}$ and $\mu_{m-}$.
The sum and difference of two squared contingency tables can be obtained by performing SVD for each case, without the need to construct a block matrix.
However, using this approach allows us to achieve optimal visualization by relying solely on the SVD of a single skew-symmetric matrix. 
Additionally, this method can be extended to more than two matched matrices while maintaining the skew-symmetric structure.

\subsection{Brief Example}\label{sec7.2}
Consider Table \ref{GSS} as a brief analysis example.
Table \ref{GSS} is taken from \cite{agresti1993computing}, with the original data obtained from the 1989 General Social Survey conducted by the National Opinion Research Center at the University of Chicago.
Subjects in the sample group were asked their opinion on (I) early teens (age 14-16) having sex relations before marriage, (II) a man and a woman having sex relations before marriage.
The response scales of premarital and extramarital sex were (1) always wrong, (2) almost always wrong, (3) wrong only sometimes, and (4) not wrong at all.
The analysis of the departures from symmetry in Table \ref{GSS} refers to exploring the degree of disagreement between views on premarital and extramarital sex.
Table \ref{SVD} shows the results of the analysis on the sum and difference in opinions between early teens and before marriage.

\begin{table}[htp]
\caption{Opinions about teenage sex, premarital sex, and extramarital sex from 1989 General Social Survey, with categories: (1) always wrong; (2) almost always wrong; (3) wrong only sometimes; (4) not wrong}
\centering
\label{GSS}
\begin{tabular}{rccccc}
\hline
 & \multicolumn{4}{c}{Extramarital sex} &  \\ \cline{2-5}
Premarital sex & (1) & (2) & (3) & (4) & Total  \\ \hline
\multicolumn{2}{l}{(I) Early teens} & \\
%\multicolumn{6}{l}{early teens (age 14-16) having sex relations before marriage} \\
(1) & 140 & 1 & 0 & 0 & 141  \\
(2) & 30 & 3 & 1 & 0 & 34 \\
(3) & 66 & 4 & 2 & 0 & 72 \\
(4) & 83 & 15 & 10 & 1 & 109 \\
Total & 319 & 23 & 13 & 1 & 356\\ 
\multicolumn{2}{l}{(II) Before marriage} & \\
%\multicolumn{6}{l}{a man and a woman having sex relations before marriage} \\
(1) & 3 & 1 & 0 & 0 & 4  \\
(2) & 3 & 1 & 1 & 0 & 5 \\
(3) & 15 & 8 & 0 & 0 & 23 \\
(4) & 23 & 8 & 7 & 0 & 38 \\
Total & 44 & 18 & 8 & 0 & 70 \\  \hline
\end{tabular}
\end{table}

\begin{figure}[htbp]
\begin{tabular}{ccc}
\begin{minipage}{.47\textwidth}
\centering
\includegraphics[width=1\linewidth]{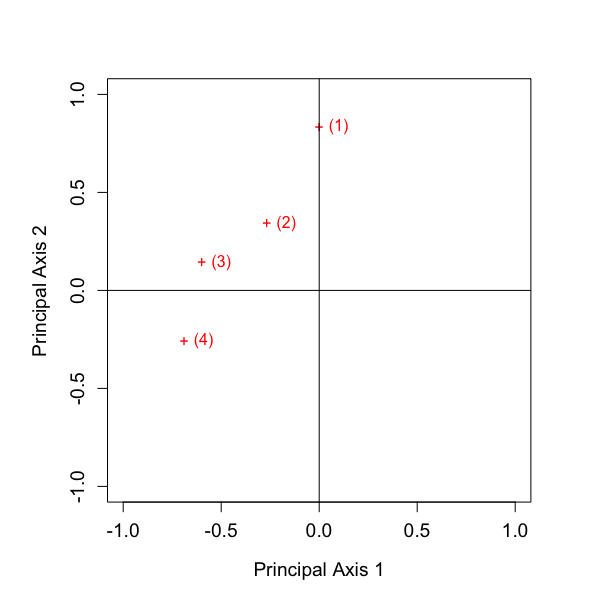}
\caption{Visualization of the sum component of the response scales}
\label{f9}
\end{minipage}
\begin{minipage}{.47\textwidth}
\centering
\includegraphics[width=1\linewidth]{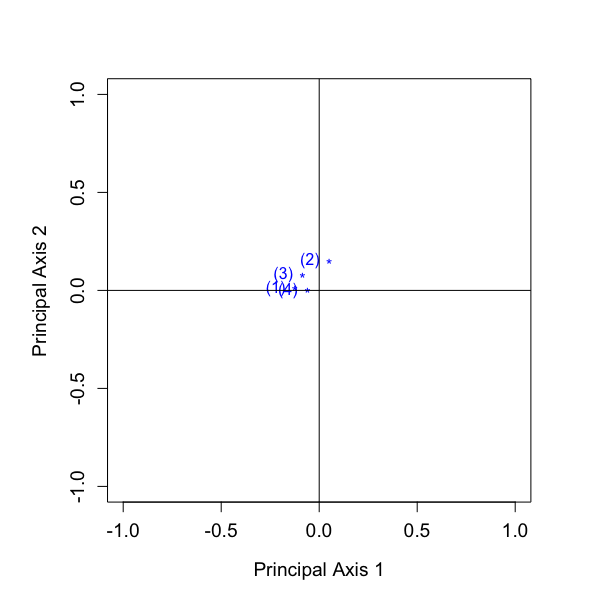}
\caption{Visualization of the difference component of the response scales}
\label{f10}
\end{minipage}
\end{tabular}
\end{figure}

The results of the SVD of the block matrix and principal coordinates without setting a metric are given in Table \ref{SVD}.
It should be noted that the singular values $\mu_{m+}$ and $\mu_{m-}$ ($m=1,\dots, 4$) of the skew-symmetric matrix related to the sum and difference components are described as follows:
\begin{align*}
\text{sum : }& \mu_{1+} = \mu_{2+} = 1.344, \; \mu_{3+} = \mu_{4+} = 0.032, \\
\text{difference : }& \mu_{1-} = \mu_{2-} = 0.239, \; \mu_{3-} = \mu_{4-} = 0.055 .
\end{align*}
From these values, dimensions 1, 2, 7, and 8 correspond to the sum component, and dimensions 3, 4, 5, and 6 correspond to the difference.
Therefore, the two CA plots of dimensions 1 and 2 together and dimensions 3 and 4 together visualize the sum and difference components in Figs \ref{f9} and \ref{f10}.
Although the consideration of a metric is usually required for constructing principal coordinates, in the current analysis, we employed the first block of the principal coordinates derived from the left singular vectors without applying any specific metric.
The primary point to highlight in Figs \ref{f9} and \ref{f10} is the differing interpretations of the origin.
Fig \ref{f9} shows the overall opinions for two sample groups, with points near the origin indicating that there was no divergence in opinion between the two groups. 
Conversely, Fig \ref{f10} presents the differences between the two sample groups, where points near the origin reflect that the two groups held similar views.

When considering the placement of the categories in the two CA plots, it becomes apparent that in Fig \ref{f9}, all categories are situated far from the origin, while in Fig \ref{f10}, they are located near the origin. 
These results suggest that irrespective of whether the group is ``early teen'' or ``before marriage'', there is disagreement in attitudes toward premarital and extramarital sex but the opinions of the two groups remain consistent across all response scales.
This consideration might be apparent even from simply reviewing Table \ref{GSS}.
However, the fact that we can draw the same insight despite a five-fold difference in sample sizes between tables of (I) and (II) is a clear advantage of our approach.

\begin{table}[htp]
\caption{SVD of $8\times 8$ block matrix and principal coordinates without a metric}
\centering
\label{SVD}
\begin{tabular}{lrrrrrrrr}
\hline
Dimension & \multicolumn{1}{c}{1} & \multicolumn{1}{c}{2} & \multicolumn{1}{c}{3} & \multicolumn{1}{c}{4} & \multicolumn{1}{c}{5} & \multicolumn{1}{c}{6} & \multicolumn{1}{c}{7} & \multicolumn{1}{c}{8} \\ \hline
\multicolumn{4}{l}{Singular values} \\
 & $1.344$ & $1.344$ & $0.239$ & $0.239$ & $0.055$ & $0.055$ & $0.032$ & $0.032$ \\ \\

\multicolumn{4}{l}{Right singular vectors} &  \\ 
First block 
 & $0.000$ & $0.622$ & $-0.518$ & $0.035$ & $-0.480$ & $0.023$ & $-0.336$ & $0.003$  \\
 & $-0.199$ & $0.255$ & $0.215$ & $0.623$ & $-0.195$ & $-0.165$ & $0.467$ & $-0.421$  \\
 & $-0.445$ & $0.107$ & $-0.351$ & $0.332$ & $0.417$ & $0.305$ & $0.202$ & $0.499$  \\
 & $-0.512$ & $-0.192$ & $-0.251$ & $-0.001$ & $0.241$ & $-0.616$ & $-0.357$ & $-0.271$  \\ \\
Second block 
 & $0.000$ & $0.622$ & $0.518$ & $-0.035$ & $0.480$ & $-0.023$ & $-0.336$ & $0.003$  \\
 & $-0.199$ & $0.255$ & $-0.215$ & $-0.623$ & $0.195$ & $0.165$ & $0.467$ & $-0.421$  \\
 & $-0.445$ & $0.107$ & $0.351$ & $-0.332$ & $-0.417$ & $-0.305$ & $0.202$ & $0.499$  \\
 & $-0.512$ & $-0.192$ & $0.251$ & $0.001$ & $-0.241$ & $0.616$ & $-0.357$ & $-0.271$  \\ \\
 
\multicolumn{4}{l}{Left singular vectors} &  \\ 
First block
 & $-0.622$ & $0.000$ & $0.035$ & $0.518$ & $-0.023$ & $-0.480$ & $-0.003$ & $-0.336$  \\
 & $-0.255$ & $-0.199$ & $0.623$ & $-0.215$ & $0.165$ & $-0.195$ & $0.421$ & $0.467$  \\
 & $-0.107$ & $-0.445$ & $0.332$ & $0.351$ & $-0.305$ & $0.417$ & $-0.499$ & $0.202$  \\
 & $0.192$ & $-0.512$ & $-0.001$ & $0.251$ & $0.616$ & $0.241$ & $0.271$ & $-0.357$  \\ \\
Second block
 & $-0.622$ & $0.000$ & $-0.035$ & $-0.518$ & $0.023$ & $0.480$ & $-0.003$ & $-0.336$  \\
 & $-0.255$ & $-0.199$ & $-0.623$ & $0.215$ & $-0.165$ & $0.195$ & $0.421$ & $0.467$  \\
 & $-0.107$ & $-0.445$ & $-0.332$ & $-0.351$ & $0.305$ & $-0.417$ & $-0.499$ & $0.202$  \\
 & $0.192$ & $-0.512$ & $0.001$ & $-0.251$ & $-0.616$ & $-0.241$ & $0.271$ & $-0.357$  \\ \\
 
\multicolumn{8}{l}{Principal coordinate by right singular vectors} &  \\ 
First block
 & $0.000$ & $0.836$ & $-0.124$ & $0.008$ & $-0.026$ & $0.001$ & $-0.011$ & $0.000$  \\
 & $-0.267$ & $0.342$ & $0.051$ & $0.149$ & $-0.011$ & $-0.009$ & $0.015$ & $-0.014$  \\
 & $-0.598$ & $0.144$ & $-0.084$ & $0.079$ & $0.023$ & $0.017$ & $0.007$ & $0.016$  \\
 & $-0.688$ & $-0.258$ & $-0.060$ & $0.000$ & $0.013$ & $-0.034$ & $-0.012$ & $-0.009$  \\ \\
Second block
 & $0.000$ & $0.836$ & $0.124$ & $-0.008$ & $0.026$ & $-0.001$ & $-0.011$ & $0.000$  \\
 & $-0.267$ & $0.342$ & $-0.051$ & $-0.149$ & $0.011$ & $0.009$ & $0.015$ & $-0.014$  \\
 & $-0.598$ & $0.144$ & $0.084$ & $-0.079$ & $-0.023$ & $-0.017$ & $0.007$ & $0.016$  \\
 & $-0.688$ & $-0.258$ & $0.060$ & $0.000$ & $-0.013$ & $0.034$ & $-0.012$ & $-0.009$  \\ \\
 
\multicolumn{8}{l}{Principal coordinate by left singular vectors} &  \\ 
First block
 & $-0.836$ & $0.000$ & $0.008$ & $0.124$ & $-0.001$ & $-0.026$ & $0.000$ & $-0.011$  \\
 & $-0.342$ & $-0.267$ & $0.149$ & $-0.051$ & $0.009$ & $-0.011$ & $0.014$ & $0.015$  \\
 & $-0.144$ & $-0.598$ & $0.079$ & $0.084$ & $-0.017$ & $0.023$ & $-0.016$ & $0.007$  \\
 & $0.258$ & $-0.688$ & $0.000$ & $0.060$ & $0.034$ & $0.013$ & $0.009$ & $-0.012$  \\ \\
Second block
 & $-0.836$ & $0.000$ & $-0.008$ & $-0.124$ & $0.001$ & $0.026$ & $0.000$ & $-0.011$  \\
 & $-0.342$ & $-0.267$ & $-0.149$ & $0.051$ & $-0.009$ & $0.011$ & $0.014$ & $0.015$  \\
 & $-0.144$ & $-0.598$ & $-0.079$ & $-0.084$ & $0.017$ & $-0.023$ & $-0.016$ & $0.007$  \\
 & $0.258$ & $-0.688$ & $0.000$ & $-0.060$ & $-0.034$ & $-0.013$ & $0.009$ & $-0.012$  \\ \hline

\end{tabular}
\end{table}

%%%%%%%%%%%%%%%%%%%%%%%%%%%%%%%%%%%%%%%%%%%%%%%%%%%%%%%%%%%%%%%%%%%%%%%%%%%%%%
\section{Conclusion}\label{sec8}
The evaluation of departures from symmetry in square contingency tables has been proposed by \cite{beh2022visualising, beh2024correspondence} through the use of generalized statistics based on power-divergence, including Bowker’s statistics, in the context of simple CA.
This paper proposes a new methodological method of simple CA based not on test statistics, but on a power-divergence measure, which serves as a measure of asymmetry.
Our approach ensures that visualizations based on well-known and interpretable divergences, such as Pearson divergence, KL divergence, and Hellinger divergence, are provided.
While both our method and previous approaches provide visualizations based on the sufficiently generalized divergence, a key distinction lies in the scaling.
Previous studies employed scaling dependent on the sample size of the observed contingency table, whereas our proposed method is independent of such scaling.
Since the measure is suitable for comparing degrees of departure across multiple contingency tables, the analysis presented in Section \ref{sec7} is feasible.
Another advantage of our proposed method is that the degrees of departure can be expressed using a skew-symmetric matrix, regardless of the parameter settings.
In the SVD of the skew-symmetric matrix, singular values are obtained in descending pairs, allowing the CA plot to be constructed in the first two dimensions, which share equal contribution ratios.
One advantage of this type of CA plot is that the interpretability of the principal coordinate plots for each category remains unaffected, even when rotations or reflections are applied.
Additionally, the row and column coordinates derived from the skew-symmetric matrix exhibit the property of matching when rotated about the origin, which ensures that the distances between the origins of identical categories in the row and column variables remain equal.
In the analysis of symmetry, this consistency of coordinates is critical for evaluating the symmetric relationships between categories.
If this consistency is lost, the interpretation of the CA plot becomes challenging, potentially compromising its reliability.
Therefore, the fact that the consistency is maintained independently of parameter settings is an important feature for analyzing symmetry.
We believe that these could provide new insights into symmetry.

\backmatter

\bmhead{Acknowledgments}
This work was supported by JST SPRING, Grant Number JPMJSP2151.
Additionally, this work was supported by the Research Institute for Mathematical Sciences, an International Joint Usage/Research Center located in Kyoto University.

\section*{Declarations}
\bmhead{Funding}
This work was supported by JSPS Grant-in-Aid for Scientific Research (C) Number JP20K03756.
\bmhead{Conflict of interest}
Not applicable.
\bmhead{Ethics approval}
Not applicable.
\bmhead{Consent to participate}
Not applicable.
\bmhead{Consent for publication}
All authors have read and agreed to the published version of the manuscript.
\bmhead{Data availability}
Not applicable.
\bmhead{Materials availability}
Not applicable.
\bmhead{Code availability}
Not applicable.
\bmhead{Authors' contributions}
These authors contributed equally to this work.

\appendix

\section{Proof of the confidence region}
\label{app:CR}
Note that the confidence region was derived based on the proposal by \cite{beh2010elliptical}.
\begin{proof}%[Proof of the confidence region for each category]
Let $a_{im}$ be the ($i, m$)th cell element ($i = 1, \dots, R; m = 1, \dots, M$) for $R \times M$ orthogonal matrix $A$ with left singular vector.
From the matrix $A$,
\begin{align*}
\sum^M_{m=1}a^2_{im} &= 1, 
\end{align*}
so that
\begin{align*}
a^2_{i1} + a^2_{i2} = 1-\sum^M_{m=3}a^2_{im}
\end{align*}
holds.
Considering the definition of the coordinates of the $i$th row category, it can also be expressed as follows
\begin{align*}
\frac{f^2_{i1}}{d^2_i\mu^2_1} + \frac{f^2_{i2}}{d^2_i\mu^2_2} = 1-\sum^M_{m=3}a^2_{im},
\end{align*}
where $f_{i1}$ and $f_{i2}$ are the ($i, 1$) and ($i, 2$)th elements of the matrix $F$ of row coordinates, and $d_i$ is the $(i, i)$th element of the diagonal matrix $D^{-1/2}$ giving the metric.
Thus, 
\begin{align*}
\frac{f^2_{i1}}{d^2_i\mu^2_1\left(1-\sum^M_{m=3}a^2_{im} \right)} + \frac{f^2_{i2}}{d^2_i\mu^2_2\left(1-\sum^M_{m=3}a^2_{im} \right)} = 1,
\end{align*}
is obtained.
This equation shows an ellipse centered at ($f_{i1}$, $f_{i2}$) whose major and minor radii are 
\begin{align*}
x_i = d_i \mu_1 \sqrt{1-\sum^M_{m=3}a^2_{im}}
\end{align*}
and 
\begin{align*}
y_i = d_i \mu_2 \sqrt{1-\sum^M_{m=3}a^2_{im}},
\end{align*}
respectively, on the two-dimensional CA plot.
Similarly, we can obtain an elliptic formula for the $j$th column category.

In order to construct a $100(1-\alpha)\%$ confidence region for the $i$th row category, consider the singular value $\mu^2_m$ of dimension $m$ and the total inertia expressed by $\Phi^{(\lambda)}$.
Assume that the upper $\alpha \%$ point of a chi-square distribution with $R(R-1)/2$ degrees of freedom is $\chi^2_{\alpha}$ and the singular value at $\chi^2_{\alpha}$ is $\hat{\mu}^2_{m(\alpha)}$.
Then, since a power-divergence statistics 
\begin{align*}
2nI^{(\lambda)}\left(\{\hat{p}^{*}_{ij};\hat{p}^{s}_{ij}\}\right) = \frac{2n}{\lambda(\lambda+1)} \mathop{\sum \sum}_{i \neq j}\hat{p}^{*}_{ij}\left[ \left( \frac{\hat{p}^{*}_{ij}}{\hat{p}^{s}_{ij}}\right)^{\lambda} - 1 \right]
\end{align*}
follows the distribution, it follows that 
\begin{align*}
\frac{\mu^2_m}{\hat{\Phi}^{(\lambda)}} &= \frac{\hat{\mu}^2_{m(\alpha)}}{\frac{\lambda(\lambda+1)}{2n\hat{\delta}(2^\lambda-1)}\chi^2_{\alpha}},
\end{align*}
and 
\begin{align*}
\hat{\mu}_{m(\alpha)} = \mu_m \sqrt{\frac{\chi^2_{\alpha}}{\hat{\Phi}^{(\lambda)}/\frac{\lambda(\lambda+1)}{2n\hat{\delta}(2^\lambda-1)}}}
\end{align*}
is obtained.
Therefore, by replacing $\mu_1$ and $\mu_2$ in $x_i$ and $y_i$ with $\hat{\mu}_1$ and $\hat{\mu}_2$, the $100(1-\alpha)\%$ confidence region for the $i$th row category around ($f_{i1}$, $f_{i2}$) is given by
\begin{align*}
\frac{(x-f_{i1})^2}{x^2_{i(\alpha)}} + \frac{(y-f_{i2})^2}{y^2_{i(\alpha)}} = 1,
\end{align*}
where
\begin{align*}
x_{i(\alpha)} &= \left(\frac{\hat{p}_{i\cdot}+\hat{p}_{\cdot i}}{2}\right)^{-1/2}\mu_1 \sqrt{\frac{\chi^2_\alpha}{\hat{\Phi}^{(\lambda)}/\frac{\lambda(\lambda+1)}{2n\hat{\delta}(2^\lambda-1)}} \left(1-\sum^M_{m=3}a^2_{im}\right) }, \\
y_{i(\alpha)} &= \left(\frac{\hat{p}_{i\cdot}+\hat{p}_{\cdot i}}{2}\right)^{-1/2}\mu_2 \sqrt{\frac{\chi^2_\alpha}{\hat{\Phi}^{(\lambda)}/\frac{\lambda(\lambda+1)}{2n\hat{\delta}(2^\lambda-1)}} \left(1-\sum^M_{m=3}a^2_{im}\right)}.
\end{align*}
Note that $M=2$ cannot be defined, in which case please refer to the proposed method such as \cite{beh2001confidence}.

\vskip-\lastskip
\end{proof}

\bibliography{sn-bibliography}% common bib file


\end{document}